\documentclass[pre,nofootinbib,twocolumn,showpacs,preprintnumbers,amsmath,amssymb,floatfix,superscriptaddress]{revtex4}
\usepackage{amsmath}
\usepackage{amsfonts}
\usepackage{dcolumn}                    	     
\usepackage{amssymb}
\usepackage{graphicx,graphics,wrapfig,rotating}     	
\usepackage[german,english]{babel}	
\usepackage{bm,fancybox}                	     

\begin{document}

\title{Distribution of resonances in the quantum open baker map}

\author{Juan M. Pedrosa}
\affiliation{Departamento de F\'\i sica, CNEA, Av. Libertador 8250, (C1429BNP) Buenos Aires, Argentina}
\author{Gabriel G. Carlo} 
\affiliation{Departamento de F\'\i sica, CNEA, Av. Libertador 8250, (C1429BNP) Buenos Aires, Argentina}
\author{Diego A. Wisniacki} 
\affiliation{Departamento de F\'\i sica, FCEyN, UBA, Pabell\'on 1 Ciudad 
Universitaria, C1428EGA Buenos Aires, Argentina}
\author{Leonardo Ermann} 
\affiliation{Departamento de F\'\i sica, CNEA, Av. Libertador 8250, (C1429BNP) Buenos Aires, Argentina}
\affiliation{Departamento de F\'\i sica, FCEyN, UBA, Pabell\'on 1 Ciudad 
Universitaria, C1428EGA Buenos Aires, Argentina}

\date{\today}


\begin{abstract}

We study relevant features of the spectrum of the quantum open baker map. The opening 
consists of a cut along the momentum $p$ direction of the 2-torus phase space, modelling 
an open chaotic cavity. We study briefly the classical forward trapped set and analyze the corresponding 
quantum nonunitary evolution operator. The distribution of eigenvalues depends strongly on the 
location of the escape region with respect to the central discontinuity of this map. This introduces 
new ingredients to the association among the classical escape and quantum decay rates. Finally, 
we could verify that the validity of the fractal Weyl law holds in all cases. 

\end{abstract}

\maketitle

\section{Introduction}

Recently, there has been an upsurge of the interest in open quantum systems, whose properties are 
still less known compared to those of the closed ones. Besides their general fundamental importance, 
they are also of the utmost relevance in very active fields. We can mention a few of them, such as 
the study of the quantum to classical correspondence \cite{Quantum2Classical}, 
quantum dots \cite{QuantumDots}, microlasers having chaotic resonant cavities 
\cite{Microlasers,Wiersig,Stone} and chaotic scattering \cite{Gaspard,Jung}. 
These systems are characterized by a nonunitary quantum evolution. The corresponding operators 
have a set of right and left decaying nonorthogonal eigenfunctions associated to them. Their complex eigenvalues 
$z_i$ (also referred to as resonances) fall inside the unit circle when represented in the complex 
plane, i.e. $\nu_i^{2}=|z_{i}|^{2}=\exp(-\Gamma_{i})\leq1$. The exponent $\Gamma_{i}\geq0$ is the usually 
called decay rate.

The classical phase space of open chaotic systems is characterized by fractal sets associated with 
trajectories that remain trapped for infinite times. Those orbits that stay forever in the future define 
what it is called the forward trapped set, and those that stay forever in the past define the 
backward one. An initial classical probability uniformly distributed 
in the phase space decays at an exponential rate. This allows 
to define the so-called classical escape rate $\gamma_{cl}$. The intersection 
of both sets, that is, the set of trajectories which do not 
escape to infinity either in the past or in the future is called the repeller. 

Regarding the eigenstates, we can distinguish between short-lived and long-lived ones. The former ($\Gamma_{i} \gg1$) 
are associated with the trajectories that escape from the system before the Ehrenfest time, while the latter 
($\Gamma_{i}= \mathcal O(1)$) are related to the classical trapped sets, thus carrying the most relevant 
classical information. One of the most important properties of open quantum systems is the conjectured fractal Weyl 
law. This law relates the mean density of resonances, $\hbar$ and the structure of the classical phase space. It 
predicts that the number of long-lived states goes as $N_{\gamma}\sim\hbar^{-(d-1)}$, where $d$ is a fractal dimension 
of the classical strange repeller. This law has been checked for a three disk system \cite{Lu} and some quantum maps 
\cite{Shepelyansky,Schomerus,Nonnenmacher1,Nonnenmacher2,Keating,Wisniacki}, and it is still being tested. 
But much less is known about the distribution of the resonances. There are some results obtained for random 
matrix models \cite{RMT}. Also, a scaling property has been numerically verified for the open kicked 
rotator \cite{Shepelyansky}. Essentially, the 
classical escape rates of this system determine the quantum decay rates associated with the long-lived 
eigenstates. However, there are no analytical results 
for the semiclassical limiting distribution. We shed new light into this open problem by concentrating on the spectral 
behavior of the most simple models of open chaotic dynamics, i.e. open piecewise linear maps. 
Discontinuities are an essential part of these systems, being responsible for their chaoticity. 
Then it is very interesting to study their influence on the spectral behavior.

In this work we focus on the quantum open baker map, which is a chaotic transformation of the unit square 
(2-torus) phase space. This is a paradigmatic model in classical and quantum chaos and also in statistical mechanics 
\cite{Saraceno1,Gaspard}. Its relevance both, in fundamental studies and in applications to a wide range of areas, 
is difficult to overestimate. As such, a deep knowledge of its spectral features is very important. We have investigated 
the behavior of the distribution of its eigenvalues as a function of the location of the escape region 
in phase space. We have found that the central discontinuity of this map plays a 
crucial role in the behavior of the spectrum. The quantum effects can be related to the classical behavior, which we 
study very briefly in order to support our explanations. But there are also 
important features of purely quantum character. In fact, the link between the classical  
escape and the quantum decay rates is more subtle for openings that overlap with the central discontinuity 
of our map than for those which do not. The shortest periodic orbits become relevant in the overlapping 
cases and based on this we provide a conjecture in order to explain our finding.

This paper is structured as follows: in Section II we describe 
the model, giving a short introduction to the classical dynamics and the quantization method used. 
In Section III we first study some aspects of the classical dynamics that help us to understand the 
spectral behavior. Then we show the results for the distribution of eigenvalues and present a 
conjecture explaining them. We also verify the validity of the fractal Weyl law for 
all cases under investigation. Finally, in Section IV we draw the conclusions. 

\section{The classical and quantum open baker map}

In this Section we introduce the main features of two dimensional torus open maps and define 
our system. We use a very simple method to model dissipation as it occurs in scattering 
through a cavity, for instance. We assume that all the classical initial conditions that are mapped 
inside the area of phase space corresponding to the opening leave the system. Thus, the open map is 
defined on a subset of the 2-torus. If we choose an opening that represents a fraction $M/N$ 
of the total area in phase space, then the open quantum map corresponds to a nonunitary matrix 
$B^o=BP$. In this expression, $P$ is a projector onto the complement of the opening, 
and the operator $B$ corresponds to the closed quantum map, a unitary matrix acting on a Hilbert 
space of dimension $N=1/2\pi\hbar$.

We study the open baker map with different opening sizes along the $p$ direction. 
The closed classical transformation is defined in the 2-torus 
$\mathcal T^{2}=[0,1)$ x $[0,1)$ by
\begin{equation}
\mathcal B(q,p)=\left\{
  \begin{array}{lc}
  (2q,p/2) & \mbox{if } 0\leq q<1/2 \\
  (2q-1,(p+1)/2) & \mbox{if } 1/2\leq q<1\\
  \end{array}\right.
\label{classicalbaker}
\end{equation}

This transformation is an area-preserving, uniformly hyperbolic, piecewise-linear and invertible map with Lyapunov 
exponent $\lambda=\ln{2}$. Geometrically, the map stretches the unit square by a factor of two in the $q$ direction, 
squeezes it by the same factor in the $p$ direction, and then stacks the right half onto the left one. 
The opening is performed by eliminating from the evolution those initial conditions falling inside a 
rectangle of width $\Delta q$, centered at $q_c$ and extending along the whole $p$ axis in phase space.
The dynamics of the open baker map has been previously studied \cite{Dorfman,Alligood}. However, the role played 
by discontinuities (specially the one along the line $q=1/2$) has not received much attention. Moreover, we 
do not know of any study regarding its effects on the spectral behavior of the quantum version. 

Following the quantization process described in \cite{Saraceno1,Saraceno2}, in an even $N$-dimensional Hilbert space, 
the quantum baker map is defined in terms of the discrete Fourier transform in 
the position representation as
\begin{equation}\label{quantumbaker}
 B_{N}=G_{N}^{-1} \left(\begin{array}{cc}
  G_{N/2} & 0 \\
  0 & G_{N/2}\\
  \end{array} \right),
\end{equation}
with $$(G_{N})_{jk}=\dfrac{1}{\sqrt{N}} \exp\{-2\pi i(j+1/2)(k+1/2)/N\}.$$

$B_{N}$ is a unitary matrix and represents the quantum dynamics of the closed baker map. 
A $\Delta q$ wide cut is made along the $p$ direction by means of the projector operator $P$ on its complement. 
Finally, the corresponding quantum dynamics for the open baker map is given by the nonunitary 
matrix $B^o_{N}=B_{N}P$. It is worth mentioning that we have chosen to open the baker map in 
a different way than in \cite{Nonnenmacher1,Nonnenmacher2}. This allows us to vary the location and width of the 
escape region.

\section{The eigenvalues: the role of discontinuities}

\subsection{The classical repeller}

We first study some features of the classical phase space of our system. 
This is solely intended to understand which 
are the main classical ingredients that play a significant part quantum mechanically. 
These ingredients will help in the explanation of the quantum behavior, which 
is the main interest of this work. For that purpose, 
we have calculated the escape rate for different locations and sizes 
of the opening. By means of the area of the forward trapped sets as 
a function of the number of the iterations of the map $A_{fw}(t)$, the classical 
escape rate can be easily calculated as $\gamma_{cl}=-\ln{A_{fw}(t)/t}$.  
Then, the information dimension $d_I$ of the corresponding repeller can be determined 
through the known relationship $d_I=2-\gamma_{cl}/\lambda$ \cite{Gaspard,Lichtenberg,Ott}.

\begin{figure}[htp]
\begin{center}
\vspace{0.025\textwidth}
\includegraphics[width=7.0cm]{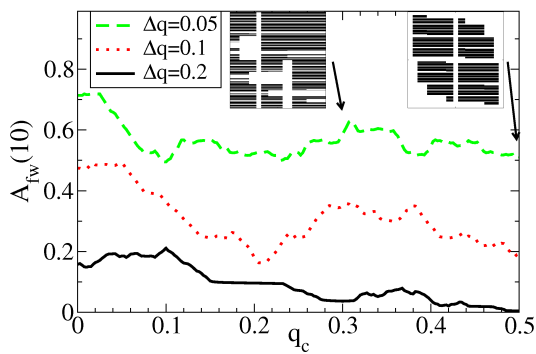}
\caption{\footnotesize(Color online): Area of the forward trapped set for the 10th 
iteration of the open baker map $A_{fw}(10)$ as a function of the center of the opening $q_c$ for different 
widths, with $\Delta q=0.2$ (black solid line), $\Delta q=0.1$ (red dotted line) and 
$\Delta q=0.05$ (green dashed line). The insets show the forward trapped sets (in black) for $t=10$ and 
$\Delta q=0.05$, for $q_c=0.3$ (left) and $q_c=0.5$ (right).}
\label{fig:repeller}
\end{center}
\end{figure}

In Fig. \ref{fig:repeller} we can see the value of $A_{fw}(10)$, i.e. the area 
of the tenth iteration of the map as a function of the position of the center of 
the opening $q_c$, for three different values of the width, $\Delta q=0.05, 0.1$ and $0.2$. 
Given that this quantity is symmetrical with respect to $q_c=0.5$ (for a given $\Delta q$), 
we only show values in the range $q_c \in [0;0.5]$.
In order to calculate these curves we have evolved initial conditions uniformly covering 
the phase space, eliminating the area corresponding to the opening at each iteration.

It can be clearly noticed that there is a common shape regardless of the size 
of the escape region. The minimum of $A_{fw}$ is at $q_c=0.5$, while the maximum is generally 
at $q_c=0$. Also, we can identify a minimum at $q_c \sim 0.25$ and 
a maximum at $q_c \sim 0.3$. This can be roughly explained by means of 
the first iterations of the opening through 
the map. In fact, the openings located at the central discontinuity typically 
do not overlap with their first iterations since there are no periodic orbits at $q=0.5$. 
As a result, initial conditions escape faster. 
This is not the case when the cut is made at the 
discontinuity at $q=0$, which includes the shortest periodic orbit ($q=0, p=0$). 
As a consequence, an opening overlapping with this discontinuity behaves much in the 
same way as one having a generic $q_c$ value. We have found that it is 
possible to use the shortest periodic orbits to give a good estimate for the 
main features of these curves (this will be explained elsewhere \cite{future}). 
In the insets we can see two examples of 
the shape of the forward trapped sets at $q_c=0.3$ and $q_c=0.5$ for $\Delta q=0.05$. These two 
positions of the escape region illustrate the two most relevant situations: in the first 
case the opening is far from the central discontinuity of the map and its first iterations, 
while in the second case it is centered on it. The maximum escape rate is 
reached for this latter and this can be easily associated with the thinner look of the 
phase space distribution. We have calculated the escape rates for $\Delta q=0.1$ with $q_c=0.3$ and 
$q_c=0.5$, resulting in $\gamma_{cl}=0.09073$ and $\gamma_{cl}=0.16488$, respectively. The 
corresponding information dimensions are $d_I=1.86910$ and $d_I=1.76213$. In the following we will 
see how this behavior translates into the quantum domain.

\subsection{The eigenvalues}

\begin{figure}[htp]
\begin{center}
\vspace{0.025\textwidth}
 \includegraphics[width=0.35\textwidth]{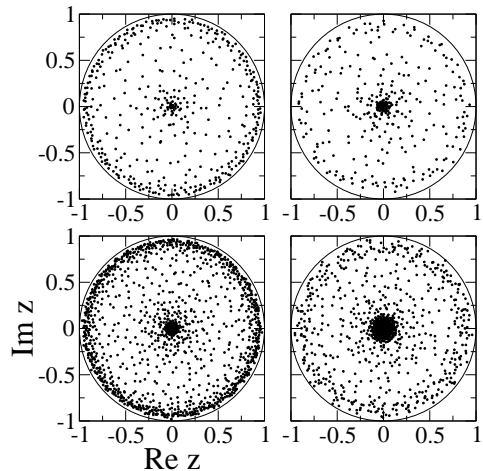}
\caption{Eigenvalues of the open baker map in the complex plane. 
In the upper panels we show the case for $N=602$ and in the lower ones for $N=2048$. 
In the left panels the opening is centered at $q_c=0.3$ and in the right ones at $q_c=0.5$; 
$\Delta q=0.1$ in all cases.}
\label{fig:perfil1}
\end{center}
\end{figure}

\begin{figure}[htp]
\begin{center}
\vspace{0.05\textwidth}
 \includegraphics[width=0.4\textwidth]{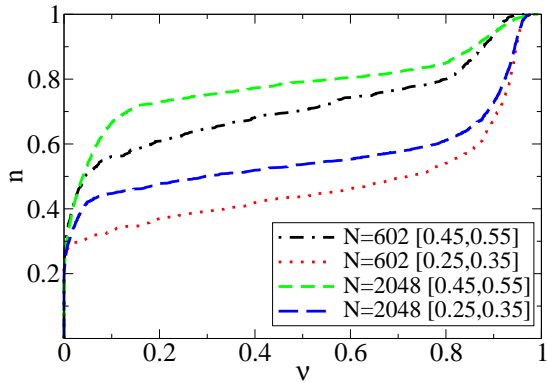}
\caption{(Color online) Cumulative number of resonances $n$ as a function of $\nu$.
The green short dashed line corresponds to $N=2048$ and the black dot dashed line 
to $N=602$, both for $q_c=0.5$; the blue long dashed line to $N=2048$ and the red 
dotted line to $N=602$, both for $q_c=0.3$.}
\label{fig:Nrvsr1}
\end{center}
\end{figure}

We now study the behavior of the distribution of the eigenvalues of the quantum evolution operator. 
We first show the eigenvalues in the 
complex plane for $\Delta q=0.1$. As can be seen in Fig. \ref{fig:perfil1}, since moduli 
are less than one ($\nu<1$) all of them fall inside the unit circle. They cluster near the origin and 
at a ring close to $\nu=1$. In the upper panels $N=602$, while in the lower ones $N=2048$. In both 
cases we display the results for $q_c=0.3$ on the left and $q_c=0.5$ on the right. In these last situations 
there is a much less dense distribution of eigenvalues at the outer ring, and an increase of density 
near the origin. This is consistent with the predictions of the fractal Weyl law, given that the 
information dimension of the classical repeller is smaller for $q_c=0.5$. It is interesting 
to mention that for an opening at $q_c=0$ the behavior is rather similar to what happens for one at
a $q_c$ far from discontinuities.
To gain further insight about the behavior of this distribution, we have calculated the normalized 
cumulative number of resonances $n=i/N$ as a function of $\nu$. In Fig. \ref{fig:Nrvsr1} 
results for the same values of $\Delta q$, $N$ and $q_c$ than in Fig. \ref{fig:perfil1} can be seen 
(see caption for details). From this Figure it is clear that there is a higher density of 
eigenvalues near $\nu=0$ for $q_c=0.5$, and a lower density near $\nu=1$ for 
$q_c=0.3$. Also, the shape of the tails ($\nu \gtrsim 0.7$) of both distributions seems to be 
different, showing a more linear behavior in the former rather than in the latter case. 

\begin{figure}[htp]
\begin{center}
\includegraphics[width=0.4\textwidth]{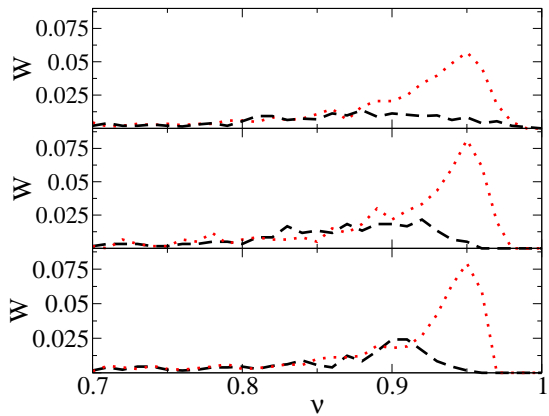}
\caption{\footnotesize(Color online) Histograms corresponding to the eigenvalue distribution $W$ as a 
function of $\nu$ for $q_c=0.3$ (red dotted lines) and $q_c=0.5$ (black dashed lines). In the 
upper panel $N=2048$, in the middle panel $N=602$, and in the lower panel $N=1782$.}
\label{fig:distr}
\end{center}
\end{figure}

In order to better evaluate the eigenvalue distribution we have constructed the histograms for $W=dn/d\nu$ 
where the bin size has been taken $\Delta \nu=0.01$. We show values for $\nu>0.7$, corresponding to the tails 
of Fig. \ref{fig:Nrvsr1}. 
To compare the distributions $W$ at different relevant cases, 
we have constructed Fig. \ref{fig:distr}. 
In these plots we have superimposed the 
cases for $q_c=0.3$ and $q_c=0.5$. In the upper panel the $N=2048$ case can be seen. This is 
an example of the special situation for the baker map when the dimension is of the form $N=2^{l}$, 
where $l$ is an integer number. In the middle panel we can see an example for $N=602$ (in this case 
$N\neq2^{l}$) where the width for $q_c=0.5$ is much greater than that for 
$q_c=0.3$. In the lower panel there is an example for $N=1782$ that shows a smaller width, 
now for the $q_c=0.5$ case. 

\begin{figure}[htp]
\begin{center}
\vspace{0.05\textwidth}
\includegraphics[width=0.4\textwidth]{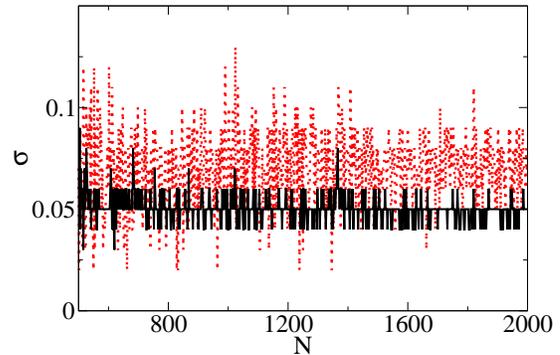}
\caption{(Color online) Width $\sigma$ of the eigenvalue distributions $W$ as a function of the 
dimension of the Hilbert space. The red dotted line corresponds to $q_c=0.5$, and the 
black solid line to $q_c=0.3$, being $\Delta q=0.1$ in both cases. Only values for $\nu>0.7$ were 
considered.}
\label{fig:distrancho}
\end{center}
\end{figure}

It is clear from Fig. \ref{fig:distr} that the width of $W$ can vary significantly with $N$. 
To have a complete picture of this behavior we have also computed the width $\sigma$ of the eigenvalue 
distributions as a function of the dimension of the Hilbert space in the interval $N \in [500;2000]$, for $\Delta q=0.1$.  
We have numerically measured the width of each histogram at half height for $\nu>0.7$. Results are shown in 
Fig. \ref{fig:distrancho}, where we have taken $q_c=0.3$ and $0.5$. We can see that the widths of the distributions 
are similar for a wide range of Hilbert space dimensions. They are generally greater for $q_c=0.5$ than 
for $q_c=0.3$ by approximately a factor of 1.5. However, there are peaks for specific 
values of $N$ where the opening contains the central discontinuity, that show both much higher and 
lower widths than for the $q_c=0.3$ case. In fact, we can identify the two peaks corresponding to 
the cases shown in the middle and lower panels of Fig. \ref{fig:distr}(i.e., $N=602$, and $N=1782$). 
It is interesting to note that these peaks seem to be present also at the high $N$ limit. 

\begin{figure}[htp]
\begin{center}
\vspace{0.05\textwidth}
\includegraphics[width=0.4\textwidth]{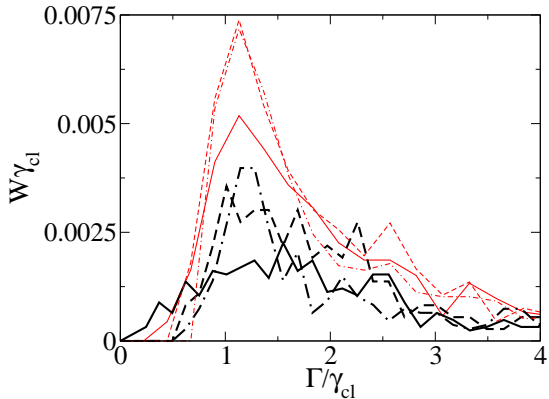}
\caption{\footnotesize(Color online) Same distributions as in Fig. \ref{fig:distr}, shown 
as a function of the decay rate and rescaled with $\gamma_{cl}$. Thin red lines correspond to $q_c=0.3$ 
and thick black ones to $q_c=0.5$. Solid lines correspond to $N=2048$, dashed to $N=602$ and 
dot-dashed to $N=1782$.}
\label{fig:distr2}
\end{center}
\end{figure}

To see more clearly the scaling property of the distributions shown in Fig. \ref{fig:distr}, we have 
represented the same data, but now as a function of the decay rate and rescaled with the classical 
escape rates $\gamma_{cl}$. We show this in Fig. \ref{fig:distr2}. 
From it, we can see that for openings with $q_c=0.3$ all distributions are almost the same, with a clear peak 
falling at $\Gamma\sim0.1\sim\gamma_{cl}$. On the other hand, 
openings overlapping with the central discontinuity show eigenvalues 
distributions that are not rescalable to the ones corresponding to the previous non-overlapping cases. Moreover, 
they are not rescalable among themselves. Fluctuations become the rule in these cases, ranging from peaks narrower 
than in the generic situations (like for $N=1782$) to distributions where no clear maximum can be found (like 
for $N=2048$ and $N=602$). This complements previous results found in the literature \cite{Shepelyansky} for 
the open kicked rotator, where this scaling turned out to be universally valid.

We could associate this behavior 
with the fact that in the $q_c=0.5$ case the shortest periodic orbits survive the dissipation process and 
have a substantial role in the localization properties of resonances. In fact, there is 
numerical evidence that individual resonance eigenstates of an open quantum system present localization 
around unstable short periodic orbits in a similar way as their closed counterparts \cite{Wisniacki}. 
This so-called scarring phenomena 
could be important enough to make almost dissapear the classical escape rate information from the quantum distributions. 
This last value is an average and now fluctuations become relevant. This includes situations 
where some particular quantization condition shrinks the 
distribution for a given $N$, as in the case of $N=1782$. In this new situation, the single 
quantum decay rates of resonances associated to given short periodic orbits could be singled out, and 
this would explain the multipeak structure of the corresponding distributions. When many orbits are 
relevant, and also longer ones, their decay rates combined could be more easily associated to a value 
that should better approximate the classical one.

\begin{figure}[htp]
\begin{center}
\vspace{0.05\textwidth}
\includegraphics[width=0.4\textwidth]{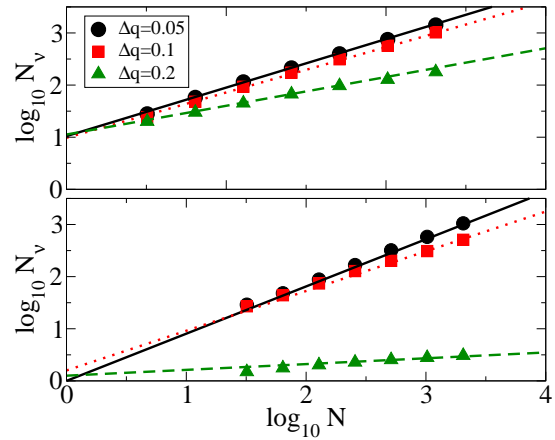}
\caption{\footnotesize(Color online) Logarithmic plot of the fraction of eigenstates $N_{\nu}$ for $\nu>0.3$, as a 
function of the dimension of the Hilbert space $N$. Lines correspond to the prediction given by the 
fractal Weyl law. Symbols correspond to the numerically calculated values. 
In the upper panel $q_c=0.3$ and in the lower panel $q_c=0.5$. In both cases 
black solid lines and circles correspond to $\Delta q=0.05$, red dotted lines and squares to $\Delta q=0.1$, 
and green dashed lines and triangles to $\Delta q=0.2$.}
\label{fig:weyl}
\end{center}
\end{figure}

Finally, we have checked whether the fractal Weyl law is verified or not by three different widths of the opening 
and for both representative $q_c$ values. The logarithmic plots of the fraction of eigenstates 
for $\nu>0.3$ as a function of the dimension of the Hilbert space can be seen in 
Fig. \ref{fig:weyl}. The lines correspond to the prediction of the fractal Weyl law 
$\log{(N_{\nu})}=\log{(N)} (d_I-1)+A$ (where $A$ is a constant). The 
symbols correspond to the numerically calculated values (see caption for more details). In all of these cases the 
slope is correctly described by the information dimension calculated from the area of the forward trapped set. We have 
adjusted the constant in order to fit the data. While in the left panel the central discontinuity
is not inside the opening and in the right one it is, we find that in both cases the agreement with the theoretical 
prediction is very good.

\section{Conclusions}

In this work we have investigated the behavior of open piecewise linear maps. 
We have found that in the open baker map the role played by the central discontinuity 
is crucial. Firstly, we have made a brief study of the properties of the classical 
forward trapped sets. We have calculated the escape rates as a function of the location 
of the opening with respect to this singularity. We have found that when a cut along the $p$ direction 
contains the discontinuity the information dimension of the repeller goes to the minimum. 
On the other hand, we have studied the behavior of the quantum map. 
We could verify that the distributions of the eigenvalues with the 
greatest classical information (i.e., with the greatest $\nu$) show a similar behavior 
for different $N$ values when the opening is located at 
a typical value of $q$ (far from the influence of the central discontinuity). 
But when this is not the case the eigenvalues behave 
in a non standard way, showing distributions that cannot be rescaled to the previous 
ones, and not even among themselves for different $N$. For the less dense fractals corresponding to a smaller 
dimension ($q_c=0.5$) the role played by the shortest periodic orbits becomes relevant. This 
supports our conjecture that they are responsible for this behavior. Also, 
fluctuations connected with the quantization rules for these trajectories 
seem to be responsible for the great differences found for different values 
of $N$. Then, the results presented in this work could be of much relevance in order to understand 
the fundamental problem of scarring phenomena present in open quantum systems. 
We think that after developing a suitable probe of localization 
for this case, an interesting study will consist of checking if the peaks 
found in the $W$ distribution correspond to short periodic orbits. Moreover, it will 
be very important to see if this behavior is also present in other types of systems, like 
open kicked maps, for example. We will investigate on this in future studies \cite{future}. 
Finally, it is remarkable that the fractal Weyl law is obeyed with great accuracy in all cases.

\begin{acknowledgments}

Partial support by ANPCyT (PICT 25373), CONICET (PIP 6137) and UBACyT 
(X237) is gratefully acknowledged.
\end{acknowledgments}

\end{document}